\documentstyle[psfig,12pt]{article}
\catcode`\@=11

\@addtoreset{equation}{section}
\def\theequation{\arabic{equation}}
\def\theequation{\thesection\arabic{equation}}

\catcode`\@=11

\@addtoreset{equation}{section}
\def\theequation{\arabic{equation}}
\def\theequation{\thesection\arabic{equation}}

\def\NPB#1#2#3{{\it Nucl.~Phys.} {\bf{B#1}} (19#2) #3}
\def\PLB#1#2#3{{\it Phys.~Lett.} {\bf{B#1}} (19#2) #3}
\def\PRD#1#2#3{{\it Phys.~Rev.} {\bf{D#1}} (19#2) #3}
\def\PRL#1#2#3{{\it Phys.~Rev.~Lett.} {\bf{#1}} (19#2) #3}

\def\@normalsize{\@setsize\normalsize{15pt}\xiipt\@xiipt
\abovedisplayskip 14pt plus3pt minus3pt%
\belowdisplayskip \abovedisplayskip
\abovedisplayshortskip  \z@ plus3pt%
\belowdisplayshortskip  7pt plus3.5pt minus0pt}
\def\small{\@setsize\small{13.6pt}\xipt\@xipt
\abovedisplayskip 13pt plus3pt minus3pt%
\belowdisplayskip \abovedisplayskip
\abovedisplayshortskip  \z@ plus3pt%
\belowdisplayshortskip  7pt plus3.5pt minus0pt
\def\@listi{\parsep 4.5pt plus 2pt minus 1pt
            \itemsep \parsep
            \topsep 9pt plus 3pt minus 3pt}}

\def\underline#1{\relax\ifmmode\@@underline#1\else
        $\@@underline{\hbox{#1}}$\relax\fi}
\@twosidetrue
\relax

\catcode`@=12

\catcode`\@=11

\def\section{\@startsection{section}{1}{\z@}{3.5ex plus 1ex minus
   .2ex}{2.3ex plus .2ex}{\large\bf}}
\def\thesection{\arabic{section}.}

\def\ps@headings{\def\@oddfoot{}\def\@evenfoot{}
\def\@oddhead{\hbox{}\hfill
        \makebox[.5\textwidth]{\raggedright\ignorespaces --\thepage{}--
        \hfill }}
\def\@evenhead{\@oddhead}
\def\subsectionmark##1{\markboth{##1}{}} }

\ps@headings

\catcode`\@=12

\relax

\def\figcap{\section*{Figure Captions\markboth
        {FIGURECAPTIONS}{FIGURECAPTIONS}}\list
        {Fig. \arabic{enumi}:\hfill}{\settowidth\labelwidth{Fig. 999:}
        \leftmargin\labelwidth
        \advance\leftmargin\labelsep\usecounter{enumi}}}
 \relax
\def\tablecap{\section*{Table Captions\markboth
        {TABLECAPTIONS}{TABLECAPTIONS}}\list
        {Table \arabic{enumi}:\hfill}{\settowidth\labelwidth{Table 999:}

        \leftmargin\labelwidth
        \advance\leftmargin\labelsep\usecounter{enumi}}}
 \relax
\def\reflist{\section*{References\markboth
        {REFLIST}{REFLIST}}\list
        {[\arabic{enumi}]\hfill}{\settowidth\labelwidth{[999]}
        \leftmargin\labelwidth
        \advance\leftmargin\labelsep\usecounter{enumi}}}
 \relax

\catcode`\@=11

\def\marginnote#1{}
\newcount\hour
\newcount\minute
\newtoks\amorpm
\hour=\time\divide\hour by60
\minute=\time{\multiply\hour by60 \global\advance\minute by-
\hour}
\edef\standardtime{{\ifnum\hour<12 \global\amorpm={am}%
    \else\global\amorpm={pm}\advance\hour by-12 \fi
    \ifnum\hour=0 \hour=12 \fi
    \number\hour:\ifnum\minute<100\fi\number\minute\the\amorpm}}
\edef\militarytime{\number\hour:\ifnum\minute<100\fi\number\minute}
\def\draftlabel#1{{\@bsphack\if@filesw {\let\thepage\relax
  \xdef\@gtempa{\write\@auxout{\string
    \newlabel{#1}{{\@currentlabel}{\thepage}}}}}\@gtempa
    \if@nobreak \ifvmode\nobreak\fi\fi\fi\@esphack}
     \gdef\@eqnlabel{#1}}
\def\@eqnlabel{}
\def\@vacuum{}
\def\draftmarginnote#1{\marginpar{\raggedright\scriptsize\tt#1}}
\def\draft{\oddsidemargin -.5truein
        \def\@oddfoot{\sl preliminary draft \hfil
        \rm\thepage\hfil\sl\today\quad\militarytime}
        \let\@evenfoot\@oddfoot \overfullrule 3pt
        \let\label=\draftlabel
        \let\marginnote=\draftmarginnote

\def\@eqnnum{(\theequation)\rlap{\kern\marginparsep\tt\@eqnlabel}%
\global\let\@eqnlabel\@vacuum}  }
\def\preprint{\twocolumn\sloppy\flushbottom\parindent 1em
        \leftmargini 2em\leftmarginv .5em\leftmarginvi .5em
        \oddsidemargin -.5in    \evensidemargin -.5in
        \columnsep 15mm \footheight 0pt
        \textwidth 250mmin      \topmargin  -.4in
        \headheight 12pt \topskip .4in
        \textheight 175mm
        \footskip 0pt

\def\@oddhead{\thepage\hfil\addtocounter{page}{1}\thepage}
        \let\@evenhead\@oddhead \def\@oddfoot{} \def\@evenfoot{}  }
\def\titlepage{\@restonecolfalse\if@twocolumn\@restonecoltrue\onecolumn
     \else \newpage \fi \thispagestyle{empty}\c@page\z@
        \def\thefootnote{\fnsymbol{footnote}} }
\def\endtitlepage{\if@restonecol\twocolumn \else  \fi
        \def\thefootnote{\arabic{footnote}}
        \setcounter{footnote}{0}}  
\catcode`@=12
\relax

\def\ps@headings{\def\@oddfoot{}\def\@evenfoot{}
\def\@oddhead{\hbox{}\hfill
        \makebox[.5\textwidth]{\raggedright\ignorespaces --\thepage{}--
        \hfill }}
\def\@evenhead{\@oddhead}
\def\subsectionmark##1{\markboth{##1}{}} }

\ps@headings

\relax

\def\firstpage#1#2#3#4#5#6{
\begin{document}
\begin{titlepage}
\nopagebreak
\title{\begin{flushright}
        \vspace*{-1.8in}
        {\small CERN-TH/98-332} \\[-10mm]
        {\small CPTH-S686.1098}\\[-10mm]
        {\small SU-ITP-98/58}\\[-10mm]
        {\small IEM-FT-184/98}\\[-10mm]
        {\small hep-ph/9810410}\\[-5mm]
\end{flushright}
\vskip -.55cm
\vfill {#3}}
\vskip -1.5cm
\author{\large #4 \\[0cm] #5}
\maketitle
\vskip -1.4 cm
\nopagebreak
\begin{abstract} {\noindent #6}
\end{abstract}
\begin{flushleft}
\rule{16.1cm}{0.2mm}\\[-3mm]
$^\dagger$ {\small On leave from: IFAE, Universitat Aut{\`o}noma de 
Barcelona, E-08193 Bellaterra, Barcelona}
\\  CERN-TH/98-332
\\[-2mm] October 1998
\end{flushleft}
\thispagestyle{empty}
\end{titlepage}}

\def\simlt{\stackrel{<}{{}_\sim}}
\def\simgt{\stackrel{>}{{}_\sim}}
\newcommand{\dal}{\raisebox{0.085cm} {\fbox{\rule{0cm}{0.07cm}\,}}}
\newcommand{\dt}{\partial_{\langle T\rangle}}
\newcommand{\dtbar}{\partial_{\langle\overline{T}\rangle}}
\newcommand{\al}{\alpha^{\prime}}
\newcommand{\mst}{M_{\scriptscriptstyle \!S}}
\newcommand{\mpl}{M_{\scriptscriptstyle \!P}}
\newcommand{\dv}{\int{\rm d}^4x\sqrt{g}}
\newcommand{\lv}{\left\langle}
\newcommand{\rv}{\right\rangle}
\newcommand{\ph}{\varphi}
\newcommand{\abar}{\overline{a}}
\newcommand{\sbar}{\,\overline{\! S}}
\newcommand{\xbar}{\,\overline{\! X}}
\newcommand{\fbar}{\,\overline{\! F}}
\newcommand{\zbar}{\overline{z}}
\newcommand{\dbar}{\,\overline{\!\partial}}
\newcommand{\tbar}{\overline{T}}
\newcommand{\taubar}{\overline{\tau}}
\newcommand{\ubar}{\overline{U}}
\newcommand{\ybar}{\overline{Y}}
\newcommand{\phb}{\overline{\varphi}}
\newcommand{\cm}{Commun.\ Math.\ Phys.~}
\newcommand{\prl}{Phys.\ Rev.\ Lett.~}
\newcommand{\pr}{Phys.\ Rev.\ D~}
\newcommand{\pl}{Phys.\ Lett.\ B~}
\newcommand{\ibar}{\overline{\imath}}
\newcommand{\jbar}{\overline{\jmath}}
\newcommand{\np}{Nucl.\ Phys.\ B~}
\newcommand{\F}{{\cal F}}
\renewcommand{\L}{{\cal L}}
\newcommand{\A}{{\cal A}}
\newcommand{\e}{{\rm e}}
\newcommand{\be}{\begin{equation}}
\newcommand{\ee}{\end{equation}}
\newcommand{\ba}{\begin{eqnarray}}
\newcommand{\ea}{\end{eqnarray}}
\newcommand{\dslash}{{\not\!\partial}}
\newcommand{\gsi}{\,\raisebox{-0.13cm}{$\stackrel{\textstyle >}
{\textstyle\sim}$}\,}
\newcommand{\lsi}{\,\raisebox{-0.13cm}{$\stackrel{\textstyle <}
{\textstyle\sim}$}\,}

\def\gappeq{\mathrel{\rlap {\raise.5ex\hbox{$>$}}
{\lower.5ex\hbox{$\sim$}}}}
\def\lappeq{\mathrel{\rlap{\raise.5ex\hbox{$<$}}
{\lower.5ex\hbox{$\sim$}}}}
\def\be{\begin{equation}}
\def\ee{\end{equation}}
\def\bea{\begin{eqnarray}}
\def\eea{\end{eqnarray}}
\def\calo{{\cal O}}
\date{}
\firstpage{3118}{IC/95/34}
{\large\bf Soft Masses in Theories with
Supersymmetry Breaking\\[-5mm] by TeV-Compactification}
{I. Antoniadis$^{\,a}$, S. Dimopoulos$^{\,b}$, A. Pomarol$^{\,c\dagger}$
and M. Quir\'os$^{\,d}$}
{\small\sl$^a$ Centre de Physique Th{\'e}orique (CNRS UMR 7644),
Ecole Polytechnique, {}F-91128 Palaiseau\\[-3mm]
\small\sl $^b$  Physics Department, Stanford University,
Stanford, California 94309, USA\\[-3mm]
\small\sl$^c$ TH-Division, CERN, CH-1211 Geneva 23,
Switzerland\\[-3mm]
\small\sl$^{d}$ Instituto de Estructura de la Materia,
CSIC, Serrano 123, E-28006 Madrid} 
{We study the sparticle
spectroscopy and electroweak breaking of theories where
supersymmetry is broken by compactification (Scherk-Schwarz
mechanism) at a TeV. The evolution of the soft terms above
the compactification scale and the resulting sparticle
spectrum are very different from those of the usual MSSM and
gauge mediated theories. This is traced to the softness of
the Scherk-Schwarz mechanism which leads to scalar sparticle
masses that are only logarithmically sensitive to the cutoff
starting at two loops. As a result, squarks and sleptons are
naturally an order of magnitude lighter than gauginos. In
addition, the mechanism is very predictive and the sparticle
spectrum depends on just two new parameters. A significant
advantage of this mechanism relative to gauge
mediation is that a Higgsino mass $\mu
\sim M_{\rm susy}$ is automatically generated when supersymmetry
is broken. Our analysis applies equally well
to theories where the cutoff is near a TeV or
$M_{P\ell}$ or some intermediate scale. We also use these observations
to show how we may obtain compactification radii which are hierarchically
larger than the fundamental cutoff scale.}

\section{Introduction}

Theories with supersymmetry softly
broken at the weak scale have been the most
popular approach to the hierarchy problem for
the last seventeen years \cite{dg}. In spite of this, the
problem of supersymmetry breaking still remains
an open question. Perhaps this is not surprising
since supersymmetry breaking is intimately connected to
the cosmological constant problem, the most
difficult problem in physics. So far there have
been two scenaria that have been suggested. One
is that there is some high scale physics of
perhaps gravitational or GUT origin leading to
the soft supersymmetry breaking terms
\cite{dg}, and is often referred
to as the ``gravity mediated scenario". The
second postulates that supersymmetry breaking originates
in a sector with which we share gauge
interactions and is called the ``gauge mediated
scenario" \cite{gr}. There have been a lot of works
in the literature dealing with the phenomenological
consequences of either gravity or gauge mediated
scenaria over the last seventeen years.

Recently a third daring possibility was suggested
\cite{a}-\cite{pq} which is based on breaking supersymmetry
by compactification (Scherk-Schwarz mechanism) \cite{ss,kp},
and involves new TeV-size spatial dimensions. The theoretical
viability of this possibility is far from obvious, since it
is embedded in theories that live in higher dimensions from
the TeV to the Planck scale. In fact, large dimensions are a
general prediction of any (known) perturbative description of
supersymmetry breaking in string theory, which necessarily
relates the breaking scale to the size of some compact
dimension(s)~\cite{ablt}.

More recently there has been a new proposal in which
the hierarchy problem is solved by lowering the
Planck scale down to a TeV \cite{aadd,l}.
This raises the possibility that the
Scherk-Schwarz mechanism for supersymmetry breaking may be
embedded in theories without any severe
ultraviolet (UV) problems. However, as we shall discuss in the
last section, in this case one has to face the problem of
the cosmological constant in the bulk which is generically much larger
than the vacuum energy on the (observable) wall.

These cautionary remarks are intended to
underline that there is no well established
framework in which the Scherk-Schwarz mechanism
for supersymmetry breaking is embedded. However,
leaving apart the cosmological constant problem, it
may be premature to dismiss the Scherk-Schwarz
mechanism as a viable possibility
for supersymmetry breaking. Moreover, as we shall
show in this paper, the spectroscopy and
experimental signatures of Scherk-Schwarz
supersymmetry breaking (SSSB) are distinct and
drastically different than either gravity or gauge mediation.

An important aspect of SSSB is that it is totally analogous
to the breaking of supersymmetry via temperature, where the
role of temperature is played by the inverse of the
compactification radius $1/R$. This implies that
supersymmetry breaking quantities are UV-insensitive. This is
a consequence of the exponential Boltzmann suppression
factors which suppress the contribution of any high energy
level to a thermodynamic quantity. In practice it means that
supersymmetry breaking physics in SSSB will only depend on
physics up to the compactification scale and will not be
sensitive to what happens beyond it. This is especially
welcome since the theory above the compactification scale is
intrinsically higher dimensional and not treatable with
standard field theory tools. The beauty of the SSSB is that
it is insensitive to the higher dimensional theory as far as
soft terms and supersymmetry breaking parameters are
concerned. Thus, in SSSB the supersymmetry breaking
parameters are under better control than supersymmetry
preserving quantities, such as gauge and Yukawa couplings.

An important corollary of this, pointed out in
ref.~\cite{add}, is that the
cosmological constant does not have quadratic
divergences or, equivalently, quadratic
sensitivity to the UV cutoff. This is analogous to
the fact that the free energy at finite temperature
is proportional to $T^4$ and has no quadratic divergences either.
Again, this follows from the exponential
suppression of high energy states'
contribution to the free energy. A testable consequence
of this behavior is the existence of light
gravitationally coupled ``moduli" with $\sim sub-millimeter$
wavelengths \cite{fkz,add}.

The aim of this paper~\footnote{The main results contained in this work have been presented at the SUSY 98 Conference, Oxford 
(11-17 July 1998)~\cite{susy98}.} is to study the
 implications of the intrinsic softness
 of the SSSB mechanism for the sparticle spectroscopy.
 We will show that the resulting sparticle
 spectrum markedly differs from the gauge or
 gravity mediated case and leads to much larger
 hierarchies of the scalar and gaugino masses
 without any fine tuning. This will be done in Sections 3 and 4.
The paper is organized as follows. In Section 2, we present
our general framework and discuss the role of extra dimensions
and the issue of gauge coupling unification.
In Section 3, we review the method
of supersymmetry breaking by SSSB compactification and
compute the one-loop corrections to the soft terms.
In Section 4, we study the breaking of the electroweak symmetry and
discuss the resulting  superparticle spectrum and its
properties.\footnote{A previous attempt to study the phenomenology of
SSSB
\cite{amq} did not take into account the extreme softness of
the soft breaking terms.}
In Section 5, we show how the softness of SSSB can
be useful in attempts to dynamically
relate the compactification and fundamental
scale (cutoff) in a hierarchical way. Finally, Section 6 contains
some concluding remarks.

\section{Large dimensions and unification}

Here, we consider a supersymmetric extension of the standard model
with an extra dimension $y$ compactified on a line interval $S^1/Z_2$,
obtained upon identification under $y\to -y$ of the points of a circle
$S^1$ of radius $R\sim$ TeV$^{-1}$. In general, the $Z_2$ discrete
symmetry acts also non trivially on the 5-dimensional (5D) fields.
These theories have two types of matter states: the bulk (untwisted)
ones, living with the gauge fields in the 5D bulk, and the
boundary (twisted) ones,
that are localized at the two fixed points of the
orbifold, $y=0$ and $y=\pi R$.

The massless spectrum has $N=1$ supersymmetry in four dimensions and
should contain the MSSM particles. The massive spectrum, however,
forms towers of Kaluza-Klein (KK) excitations for all the
fields living in the 5D bulk,
with masses $n/R$ for $n=0,1,\dots$; they fall into
supermultiplets of extended $N\ge 2$ supersymmetry.
We will distinguish two different cases:\\
(a) The KK modes are organized either in $N=4$
supermultiplets, or in $N=2$ but are falling into appropriate group
representations leading to vanishing beta-functions. For brevity, we
shall refer to this case as the $N=4$ one.\\
(b) The KK modes form just $N=2$ multiplets.\\
On the other hand, obviously, the 4D boundary fields have no KK
excitations.

Following refs.~\cite{amq,pq}, we will consider that in addition to
gauge
multiplets only the Higgs fields live in the 5D bulk, as a part of $N=2$
vector supermultiplets or hypermultiplets.
In this way, the $\mu$-problem is automatically solved, since
as we shall see in the next section, Higgsinos acquire a mass of
the order of the compactification scale by the Scherk-Schwarz mechanism
of  supersymmetry breaking. On the other hand,
quarks and leptons chiral multiplets are assumed to
be localized in the 4D boundary.

In the $N=4$ case (a), there is no contribution from the KK
states to the beta-function coefficients of the gauge
couplings. The only contribution arises from the zero modes
and the twisted states of the 4D boundary. Therefore, the
gauge couplings evolve as in the MSSM and unify at
$M_{st}\simeq 10^{16}$ GeV. In general, in these models,
additional constraints have to be imposed in order to avoid
potential growing of the Yukawa couplings. For instance, when
the quarks and leptons are also bulk fields and come from
$N=2$ hypermultiplets, this condition is automatically
satisfied, since their wave function is not renormalized. Of
course, in this case, special model building is needed to
satisfy the condition of vanishing of the 1-loop $N=2$
beta-functions. When quarks and leptons are twisted fields,
the Yukawa coupling constraints are satisfied if for
instance there are non-trivial infrared-stable fixed points
in the full theory \footnote{This possibility was pointed out
to us by Nima Arkani-Hamed.}.

In the generic $N=2$ case (b), the KK states contribute to the gauge
beta-function coefficients,
and change the logarithmic scale dependence  of the gauge coupling
to a  power-law running. This
accelerates the gauge couplings evolution and they may unify
at much lower energies \cite{ddg}.
Assuming that only the gauge and Higgs fields live in 5D,
and that supersymmetry is broken
by the Scherk-Schwarz mechanism along the orbifold compactification
at the scale $M_c\equiv 1/R$, one has at one loop level
\be
\frac{1}{\alpha_i(m_Z)}=\frac{1}{\alpha_{st}}+
\frac{b_i^{SM}}{2\pi}
\ln\frac{M_c}{m_Z}+
\frac{b_i^{MSSM}}{2\pi}
\ln\frac{M_{st}}{M_c}+
\frac{b_i^{KK}}{2\pi}\left[\frac{M_{st}}{M_c}-1
-\ln\frac{M_{st}}{M_c}\right] +\Delta_i\, ,
\label{run}
\ee
where $b^{SM}=(41/10,-19/6,-7)$, $b^{MSSM}=(66/10,1,-3)$ and
$b^{KK}=(3/5,-3,-6)$ are respectively the
beta-function coefficients of the Standard Model (SM),
MSSM and KK states, while $\Delta_i$ denote additional
string threshold corrections.

In the absence of threshold corrections, it turns out that the measured
values of the three gauge couplings at $m_Z$ lead to an approximate
unification (within $2\%$ taking into account the experimental errors)
with
\be
M_{st}\simeq 45\ {\rm TeV}\, \qquad  \alpha^{-1}_{st}\simeq 50\, ,
\label{mst}
\ee
for $M_c\simeq 1$ TeV. However, as one can see from
eq.~(\ref{run}), the gauge couplings acquire a power law
sensitivity with respect to the UV cutoff $M_{st}$. As a
result, the gauge coupling unification conditions are
extremely sensitive to string threshold corrections~\cite{Ross}. This
casts doubts on the significance of this
calculation. It shows that unification of couplings cannot be
decided by a low energy effective-theory computation; it
requires a detailed knowledge of the full UV theory.

It is important to point out the relevance of the underlying string
theory in both cases we discussed above, and mainly in the $N=4$ case. 
Unlike gauge
couplings whose quantum corrections are constrained by holomorphicity,
physical amplitudes cannot in general be reliably computed beyond
the compactification scale, as the effective field theory becomes
non-renormalizable (higher dimensional). The full string theory is then
required above $M_c$ to fix the coefficients of the higher dimensional
effective operators. However, as we will see in the next section, this
``pathology" does not hold for the effective couplings that are
generated
after supersymmetry breaking, due to the extreme softness of the SSSB
mechanism.

\section{Soft terms in Scherk-Schwarz compactification}

Theories with extra dimensions, and in particular five
dimensional theories,
allow the use of the Scherk-Schwarz (SSSB) mechanism
to break supersymmetry \cite{ss,pq}.
This consists in
imposing to the 5D fields a  different  periodicity condition
for bosons and fermions
under a $2\pi R$
translation of the extra dimension:
\begin{equation}
\Phi(x^\mu,y+2\pi R)=e^{2\pi i q_{\Phi} }\Phi(x^\mu,y)\, ,
\label{SS}
\end{equation}
where $q_{\Phi}$ is the R-symmetry charge of the field $\Phi$.
Due to the periodicity condition (\ref{SS}), the fields
are Fourier expanded as
\be
\Phi(x^\mu,y)= \sum^{\infty}_{n=-\infty}e^{iy(n+q_{\Phi})/{R}}\,
\Phi^{(n)}(x^\mu)\, ,
\label{expansion}
\ee
where $y$ is assumed to be compactified on the circle $S^1$.
Reducing the theory from 5D to 4D,
eq.~(\ref{expansion}) leads to a tower of KK states with a
fermion-boson  mass splitting inside each KK supermultiplet:
\bea
m^2_B&=&{(n+q_B)^2}\,{M^2_c}\ ,\nonumber\\
m^2_F&=&{(n+q_F)^2}\,{M^2_c}\  ,\ \ \ n=0,\pm 1,\pm 2,...\, ,
\label{splitting}
\eea
where $M_c\equiv 1/R$, and $q_B$ and $q_F$ are the charges of the
bosons and fermions, respectively.

In orbifold compactifications, the expansion (\ref{expansion})
is truncated, since only the invariant states under the orbifold group
remain in the theory.
For example in $S^1/Z_2$ compactifications, the $Z_2$
parity, $y\rightarrow -y$, acts
on the KK-states as  $\Phi^{(n)}\rightarrow \Phi^{(-n)}$.
Therefore the $Z_2$ projects the KK-tower into
$Z_2$-even ($\Phi^{(n)}_+$)
or $Z_2$-odd ($\Phi^{(n)}_-$) states:
\begin{eqnarray}
\Phi^{(n)}_+&=&\Phi^{(n)}+\Phi^{(-n)}, \qquad n=0,1,2,...\, ,\nonumber\\
\Phi^{(n)}_-&=&\Phi^{(n)}-\Phi^{(-n)}, \qquad n=1,2,...\, .
\label{evenodd}
\end{eqnarray}

Massless fields arise in the theory
only if either $q_B$ or $q_F$ are zero.
We are interested in the $q_B=0$ case, in which
only the  vector and scalar boson $n=0$ states remain in the
massless spectrum of the theory,
and can be associated with the bosonic sector of the SM.
Although vector bosons remain massless because of the gauge symmetry,
the $n=0$ complex scalar field, $\phi$, will get a mass at the one-loop
level due to the breaking of  supersymmetry by the SSSB mechanism.
Each level of KK excitations contribute to the scalar mass.
This is a crucial difference with respect to 4D theories with
softly broken supersymmetry.

The one-loop  $m^2_\phi$ induced by a tower of KK
with a mass splitting (\ref{splitting})
can be obtained from the effective potential $V(\phi)$:
\begin{equation}
m^2_\phi=\left.\frac{d\, V(\phi)}{d|\phi|^2}\right|_{\phi=0}\, ,
\label{maspot}
\end{equation}
with $V(\phi)$ given by
\begin{equation}
V(\phi)=\frac{1}{2}{\rm Tr}
\sum^{\infty}_{n=-\infty}
\int\frac{d^4 p}{(2 \pi)^4}
\ln\left[\frac{p^2+(n+q_B)^2\, M_c^2+
M^2(\phi)}{p^2+(n+q_F)^2\, M_c^2+M^2(\phi)}\right]\, ,
\end{equation}
where the trace is over the degrees of freedom of the KK tower
and ${M^2(\phi)}$ is the $\phi$-dependent mass of the KK states.
{}From eq.~(\ref{maspot}), we obtain
\begin{eqnarray}
m^2_\phi&=&
\frac{1}{2}{\rm Tr} \left.\frac{dM^2(\phi)}{d|\phi|^2}\right|_{\phi=0}
\sum^{\infty}_{n=-\infty}\Pi_n(0) \, ,\label{sum}\\
\Pi_n(0)&=&\int\frac{d^4p}{(2 \pi)^4}\left[
\frac{1}{p^2+(n+q_B)^2M^2_c}-\frac{1}{p^2+(n+q_F)^2M^2_c}\right] \, .
\label{kkcont}
\end{eqnarray}
As in finite temperature calculations, we must first
sum over the infinite tower of  KK states and then perform
the momentum integral.
Eq.~(\ref{kkcont}) thus leads to \footnote{For further details see
ref.~\cite{prep}.}
\begin{equation}
m^2_\phi=
\frac{1}{32\pi^4}
\left[\Delta m^2(q_B)-\Delta m^2(q_F)\right]
{\rm Tr}\left.\frac{dM^2(\phi)}{d|\phi|^2}\right|_{\phi=0}
\label{kkcont2}\, ,
\end{equation}
where
\begin{equation}
\Delta m^2(q)=\frac{1}{2}(Li_3(z)+Li_3(1/z))\, M^2_c\, ,
\label{lit}
\end{equation}
with $z\equiv e^{i2\pi q}$ and
$Li_n(z)$ are the polylogarithm functions $Li_n(z)=\sum^{\infty}_{k=1}
\frac{z^k}{k^n}$. For $q$ ranging from $0$ to $1/2$,
$(Li_3(z)+Li_3(1/z))/2$ goes from $1.2$ to $-0.9$.

We have just considered a theory compactified on $S^1$.
For a compactification on the $S^1/Z_2$ orbifold,
the contribution to $m^2_\phi$ from a KK-tower
must sum from $n=0$ to $\infty$ (see eq.~(\ref{evenodd})).
Nevertheless, the contribution to $m^2_\phi$
of an even field ($\Pi^+_n$) can be always added to
the one of an odd field ($\Pi^-_n$)
such that they can be considered
as  the contribution of a  single KK-tower
 where $n$ goes from $-\infty$ to $\infty$ \cite{pq}:
\begin{equation}
\sum^{\infty}_{n=0}\Pi^+_n+\sum^{\infty}_{n=1}\Pi^-_n=
\sum^{\infty}_{n=-\infty}\Pi_n\, ,
\end{equation}
where $\Pi_{\pm n}\equiv \Pi^\pm_n$.
Thus, in  $S^1/Z_2$ orbifold compactifications,
the effective number of states of the KK-tower is reduced by a  half
with respect to that in  $S^1$.

The contribution (\ref{kkcont2}) is finite and ultraviolet independent.
There are different ways to understand this result.
The simplest way
is to notice that a theory
with SSSB supersymmetry breaking keeps a clear
analogy with a theory at
finite temperature $T$
(a quantum field theory with  the time compactified)
where $M_c$ plays the role of $T$.
At finite temperature, the $T$-dependent
effective potential is not affected by the ultraviolet cutoff
due to the Boltzmann suppression of the heavy states.
Of course, the $T$-independent part of the
effective potential is ultraviolet
sensitive. In our case, however, $M_c=0$ corresponds to the
supersymmetric limit and therefore the corrections to
$V(\phi)$ are zero.
Alternatively, the insensitivity of $m^2_\phi$ to an ultraviolet
scale $\Lambda\gg M_c$ (where new physics arises)
can be probed by just putting an explicit ultraviolet cutoff
in our theory and showing that the result
(\ref{kkcont2}) is not modified.
This is in
analogy with Casimir energy calculations \cite{casimir}.
For example,
we can insert in the sum  of eq.~(\ref{sum})
a function $f(\Lambda,nM_c)$
that goes to zero for $nM_c\gg \Lambda$.
The function $f$ must be normalized such that
$f\rightarrow 1$ for $n\rightarrow 0$.
For example, we can take
$f=e^{-nM_c/\Lambda}$.
As expected, we find that
the calculation of  $m^2_\phi$  gives,
for $\Lambda\gg M_c$,
the same result (\ref{kkcont2}).

Using eq.~(\ref{kkcont2}) we can calculate
the contribution to the soft mass of any scalar field of the theory.
Let us take as an example, the model of ref.~\cite{pq} in which
after compactification on $S^1/Z_2$ a
massless scalar  $\phi$ arises from a 5D hypermultiplet.
For simplicity, let us assume $q_F=1/2$ and $q_B=0$.
The fermionic $Z_2$-even sector consists of two bispinors, a gaugino
and the partner of $\phi$, that combine with the odd sector
to form two full KK-towers.  We then have
$4$ degrees of freedom
and  ${\rm Tr}M^2(\phi)=8g^2C(\phi)|\phi|^2$ where $C(\phi)$ is the
quadratic Casimir of the scalar $\phi$ in the corresponding gauge group
[$C({\bf N})=(N^2-1)/(2N)$  for the fundamental representation of
SU(N)].
Therefore
\begin{equation}
{\rm Tr}\, \left.\frac{dM^2(\phi)}{d|\phi|^2}\right|_{\phi=0}=8g^2
C(\phi)\, .
\label{nf}
\end{equation}
By supersymmetry eq.~(\ref{nf}) must hold both for bosons and fermions.
Using eqs.~(\ref{kkcont2}) and (\ref{nf}), we obtain
\begin{equation}
m^2_\phi=\frac{7g^2C(\phi)\zeta(3)}{16\pi^4}M^2_c\simeq
5\times 10^{-3} M^2_c\, ,
\label{5Dmass}
\end{equation}
where for the numerical estimate
we have taken  $g^2C(\phi)\sim 1$ and $\zeta(3)\simeq 1.2$.
Thus, the scalar remains around an order of magnitude lighter than
the compactification scale.

Let us now consider  the fields living in the 4D boundary.
These fields do not have associated KK excitations and
are massless at tree level.
Nevertheless, if they couple to the fields
living in the 5D bulk,
the supersymmetry breaking will be transmitted from the
bulk to the boundary and, as a consequence,
the scalars living in the boundary will get
masses at the one-loop level.
Let us consider again the case
of an $S^1/Z_2$ orbifold.
The possible interactions between the bulk and the boundary fields
can be found in ref.~\cite{peskin}.
The gauge interactions couple the fields in the boundary to
the KK-towers of the gauge boson, gaugino and the auxiliary
$D$-field. At the one-loop, we find that the
boundary scalars get a mass given by (for $q_B=0$)
\begin{equation}
m^2_i=
\frac{g^2C(R_i)}{4\pi^4}
\left[\Delta m^2(0)-\Delta m^2(q_F)\right]\, ,
\label{gaugecont}
\end{equation}
where $R_i$ is the representation of the gauge group under which
the boundary field transforms, and $\Delta m^2(q)$ is given in
eq.~(\ref{lit}). The boundary field can also couple
to an $N=1$ chiral supermultiplet that consists in the KK-towers of a
complex scalar, a bispinor and the auxiliary $F$-field.
In this case we find that
the scalar field of the boundary gets a mass given by
\begin{equation}
m^2_i=
\frac{Y^2}{16\pi^4}
\left[\Delta m^2(q_{B})+\Delta m^2(2q_F-q_{B})
-2\Delta m^2(q_{F})
\right]\, ,
\label{yukawacont}
\end{equation}
where $Y$ is the  Yukawa coupling
between the bulk and boundary fields.
The first term in eq.~(\ref{yukawacont}) arises from the scalar
KK-tower, the second term from the auxiliary $F$-field KK-tower  and
the third from the bispinor KK-tower.
Again these contributions are also finite and ultraviolet independent.

{}Finally, we have calculated
the contribution of the KK-towers to a scalar trilinear coupling, $A$,
between two boundary fields, $Q$ and $U$,
and one field in the bulk.
This contribution arises from gaugino loops and gives
\begin{equation}
A=4Yg^2T^a_QT^a_U
\int\frac{d^4p}{(2 \pi)^4}\frac{1}{p^2}
\sum^{\infty}_{n=-\infty}
\frac{(n+q_F)M_c}{p^2+(n+q_F)^2M^2_c}\, ,
\end{equation}
that leads to
\begin{equation}
A=\frac{Yg^2T^a_QT^a_U}{8\pi^3}\Delta A(q_F)\, ,
\end{equation}
where
\begin{equation}
\Delta A(q_F)=i
\left[Li_2(e^{i2\pi q_{F}})-Li_2(e^{-i2\pi q_{F}})\right]\,M_c\, ,
\end{equation}
and $T^a_R$ is the generator of the  gauge group in the
representation of $R$.

Note that in contrast to the case of supersymmetric parameters, as gauge
and Yukawa couplings that exhibit generically a power dependence on the
UV
cutoff, the soft breaking parameters are insensitive to it. This is
due to the extreme softness of the SSSB mechanism which dies off
exponentially in the supersymmetric limit, in analogy to the situation
at
finite temperature. This behavior persists, as well, for all couplings
of
higher dimensional operators induced by the supersymmetry breaking. In
fact, it is easy to see that these couplings vanish in the
supersymmetric
limit and they are suppressed by powers of $M_c/M_{st}$.

Let us finally comment on higher loop effects.
In analogy with the finite temperature calculation,
we can implement some of these effects
by replacing the couplings $g$ and $Y$ in the above calculations
by the running couplings at the scale $M_c$, $g(M_c)$ and $Y(M_c)$.
The effect of the boundary fields, however, does not
have an analogy with finite temperature.
The fields in the boundary contribute at the one-loop level to
the wave function renormalization constant of the KK-excitations.
This contribution will make the KK masses to evolve logarithmically
with the renormalization scale.
As a consequence, at higher-loop orders, $M^2_c$   must
be  replaced  by the renormalized $M^2_c$ (see below eq.~(\ref{etai})).

\section{Superparticle spectrum and electroweak symmetry breaking}
\subsection*{Sspectroscopy and the LSP}
Let us apply the above calculation to our model, where gauginos
and Higgsinos in the bulk are given the same boundary
conditions, $q_F$, corresponding to a common $R$-symmetry
charge. A more general case \cite{pq} will be treated in \cite{prep}.
Since gauge and Higgs bosons live in the 5D bulk their
corresponding $n=0$ fermions, gauginos and Higgsinos,
will get masses of order $M_c$:
\begin{eqnarray}
m_\lambda&=&q_F\, M_c\, ,\\
m_{\widetilde H}&=&q_F\, M_c\, .
\end{eqnarray}

Therefore the massless states of the 5D bulk correspond to the gauge and
Higgs bosons of the SM.
Quarks and leptons superfields
reside in the 4D boundary, and then
they also remain massless at the tree-level.
Nevertheless, since supersymmetry is broken
in the bulk,
squarks and sleptons will get masses at the one-loop level
through  the gauge  and Yukawa interactions,
leaving only the fermion sector
of the SM in the massless spectrum.
Using eqs.~(\ref{gaugecont}) and (\ref{yukawacont}), we obtain the
squarks and sleptons masses:
\begin{eqnarray}
m^2_{\widetilde Q}&=&
\left(\frac{8}{6}\alpha_3+\frac{3}{4}\alpha_2+
\frac{1}{60}\alpha_1\right)\Delta m_{g}^2
+\frac{1}{2}\alpha_t\Delta m_{H}^2\, ,\\
m^2_{\widetilde U}&=&\left(\frac{8}{6}\alpha_3+
\frac{4}{15}\alpha_1\right)\Delta m_{g}^2
+\alpha_t\Delta m^2_{H}\, ,\\
m^2_{\widetilde D}&=&\left(\frac{8}{6}\alpha_3+
\frac{1}{15}\alpha_1\right)\Delta m_{g}^2\, ,\\
m^2_{\widetilde L}&=&\left(\frac{3}{4}\alpha_2+
\frac{3}{20}\alpha_1\right)\Delta m_{g}^2\label{lepton}\, ,\\
m^2_{\widetilde E}&=&\frac{3}{5}\alpha_1\Delta m^2_{g}\, ,
\end{eqnarray}
where
\begin{equation}
\Delta m_{g}^2=
\left[\Delta m^2(0)-
\Delta m^2(q_{F})\right]/\pi^3\, ,
\end{equation}
and
\begin{equation}
\Delta m_{H}^2=
\left[\Delta m^2(0)+\Delta m^2(2 q_{F})-2\Delta m^2(q_{F})
\right]/(2\pi^3)\, ,
\end{equation}
with
$\Delta m^2(q)$ given in eq.~(\ref{lit}).

The above equation gives us a very predictive spectrum for the
squarks and sleptons.
It only depends on two free parameters, $q_F$ and $M_c$.
Notice also that the above contributions are positive
as they are necessary to avoid color or charge breaking.
The ratio of masses is given by  (for $q_F=1/2$)
\begin{equation}
10\,{m_{\widetilde Q}}\,\simeq\,10\,{m_{\widetilde D}}\,\simeq\,10\,
{m_{\widetilde U}}
\,\simeq\, 25\, {m_{\widetilde L}}\,\simeq \, 40\, {m_{\widetilde
E}}\,\simeq \, m_{\lambda, \widetilde H}\, .
\end{equation}

Therefore the gauginos and Higgsinos are the heaviest
supersymmetric particles and the right-handed slepton is the
lightest (LSP) one. In an R-parity conserving theory, the
right-handed slepton will be stable and will cross the
detector leaving an ionizing track. It can be discovered by
looking at anomalous ionization energy loss, $dE/dx$, in the
tracking detector gas \cite{opal}. The actual experimental
lower bound on its mass is $82.5$ GeV \cite{opal}.
Cosmological arguments all but exclude charged, stable
particles with masses in the $\sim 100$ GeV to $\sim 10$
TeV range \cite{derujula}. A significant number of these
particles will survive annihilation with their antiparticles
and, at the time of Nucleosynthesis, combine with other
nuclei to form ``heavy'' hydrogen, helium, etc., leading to
heavy versions of these atoms today. Searches for such
anomalous isotopes put very strong limits on their fractional
abundance --as small as $10^{-30}$ for hydrogen. There is a
variety of other possible considerations showing the
implausibility of stable-charged LSPs, including interstellar
calorimetry --the thermodynamics of interstellar clouds
\cite{cohen}-- as well as neutron stars \cite{gould}. 

One way
out of this is to ensure that the charged LSP is unstable. A
simple way to accomplish this is to postulate an R-parity
breaking interaction. Indeed, if R-parity is violated, the
right-handed slepton will decay into SM leptons. In fact,
since the R-parity violating coupling is renormalizable, the
slepton is expected to decay inside the detector and can be
easily discovered \cite{aleph}. Another possibility to avoid
a right-handed slepton LSP occurs in theories with
right-handed neutrinos~\cite{k}. These must have miniscule Yukawa
coupling to account for the observed smallness of the
neutrino masses. Since the right-handed sneutrinos are
electroweak singlets and also have miniscule Yukawa couplings,
they will get tiny masses of order of the (Dirac) neutrino
masses in the sub-eV range and will be the LSP. We are not
aware whether such an LSP passes all the necessary cosmological
safety tests, but it does not seem to us to be obviously
excluded.

\subsection*{Electroweak Breaking}

To study the breaking of the electroweak symmetry, we must
analyze the soft masses of the Higgs(es).
In the MSSM, we need two  Higgs SU(2)$_L$-doublets,
$H_1$ and $H_2$, in order to give masses to all the fermions living
in the 4D boundary.
After imposing the supersymmetry breaking with the SSSB
compactification,
these two Higgses can either arise as two
(tree-level) massless
states, or as a massless and a massive state.
This depends on the  R-charges of the Higgses,
and different possibilities have been proposed in refs.~\cite{amq,pq}.
The simplest case corresponds to having a unique massless
SU(2)$_L$-doublet, $\phi$, that will be  responsible for the
electroweak symmetry breaking.
This massless field can be either associated
to one of the MSSM Higgses or to
a linear combination of the two, $\phi\equiv\cos\beta H_1+\sin\beta
H_2$.
The mixing angle $\tan\beta$ is model dependent.
For example, for the model
of ref.~\cite{pq} in which $\phi$ arises from a 5D hypermultiplet
and corresponds to a flat direction of the  D-terms,
one has $\cos\beta=\sin\beta$.
In ref.~\cite{amq} the only massless mode is the Higgs that couples
to the top-quark: this would correspond to $\sin\beta=1$.
Further scenarios will be considered in ref.~\cite{prep}.

Here we will be only interested in knowing whether
$\phi$ gets a vacuum expectation value and therefore
breaks the electroweak symmetry. For this purpose we must calculate the
quantum corrections to its mass.
Using eq.~(\ref{kkcont2}), we have
\begin{equation}
m^2_{\phi}(M_c)=\left(\frac{3}{4}\alpha_2+
\frac{3}{20}\alpha_1\right)\Delta m_{g}^2\, .
\end{equation}
This mass is positive.
Nevertheless, we  must also consider
the correction to the Higgs mass due to the
stop. This correction arises at the two-loop level but it
is important since the stops are heavier than the Higgs.
This is given by
\begin{equation}
m^2_\phi(m_Z)\simeq m^2_\phi(M_c)-\frac{3\alpha_2m^2_t}{8\pi m^2_W}
(m^2_{\widetilde Q}+m^2_{\widetilde U})\ln\frac{M^2_c}{m^2_{\widetilde
Q}}\, .
\label{stopcont}
\end{equation}
As in theories of gravity or gauge-mediation supersymmetry breaking,
this contribution turns the Higgs mass to negative values
and triggers the breaking of the electroweak symmetry.
Imposing the minimization condition
\begin{equation}
  -m^2_\phi(m_Z)=\frac{m^2_Z}{2}\cos^2 2 \beta\, ,
\label{elmin}
\end{equation}
one  can derive the value of the compactification
scale $M_c$. Taking $q_F=1/2$,
$m_t\simeq 175$ GeV and $\cos 2\beta\sim 1$, we obtain
\begin{equation}
M_c\sim 3.7\ {\rm TeV}\ ,\ \ \ \ \
m_{\widetilde Q}\sim 400\ {\rm GeV}\ ,\ \ \ \ \
m_{\widetilde E}\sim 100\ {\rm GeV}\, .
\label{spectrum}
\end{equation}
We must notice that the above prediction is
quite sensitive to the values of $m_t$ and $\alpha_3$ that have large
experimental uncertainties. We find some
 values for these parameters for which
the two terms of the RHS of eq.~(\ref{stopcont}) approximately
 cancel out,
 and consequently the supersymmetric spectrum turns to be much heavier.

\subsection*{Contrast with the MSSM and Gauge Mediation}

Sparticle spectroscopy in SSSB is strikingly different from
that of more familiar gauge mediated and MSSM. All
fermionic sparticles are at least an order of magnitude
heavier than the bosonic ones, a smoking gun for this
framework. In particular, the gauginos and higgsinos are
$\sim 40$ times heavier than the right-handed sleptons. A
consequence of this, following from the present lower limit
of $82.5$ GeV on the mass of the right-handed sleptons, is
that the compactification scale as well as the gaugino and
higgsino masses must be no less than 3 TeV. Consequently, the
KK excitations of ordinary particles would not be accessible
at LHC.

Another point of contrast is the ease with which SSSB
generates a mass $\mu$ for the higgsinos. This occurs as an
integral part of the SSSB and naturally accounts for the
equality of the $\mu$ term and the supersymmetry breaking
scales. In contrast, in gauge mediation one needs to work
hard to accomplish this task \cite{dgp}.

Finally there are some similarities. As in gauge-mediated
theories, the breaking of supersymmetry is communicated to
the squarks by the gauge interactions. Therefore the theory
does not have dangerous flavor violating interactions.

\section{Dynamical determination of the compactification radius}

In this section we determine the value of the radius by minimizing the
vacuum energy with respect to the corresponding modulus field, or
equivalently with respect to the compactification scale $M_c$. As we
mentioned already in the introduction, the Higgs contribution at the
electroweak breaking minimum (\ref{elmin}), being proportional to
$m_H^4(M_c)$, is negligible compared  to the direct contribution
computed
in ref.~\cite{add}. The latter reads (in the large radius limit):
\ba
E &=& {1\over 2}{\rm Str}\int{d^4 p\over (2\pi)^4}\ln\left\{
p^2\left(1-\gamma_T{\alpha\over 2\pi}\ln{p^2\over \Lambda^2}\right)
+M^2\right\}+\cdots\nonumber\\
&=& \sum_i\eta_i \left(1+4 \gamma_i{\alpha_i\over 2\pi}
\ln{M_c\over \Lambda}\right)M_c^4
+\cdots\, ,
\label{En}
\ea
where $\Lambda$ is the cutoff scale, and
all couplings in (\ref{En}) are considered at the scale $\Lambda$.
In this context the cutoff
$\Lambda$ is the scale at which the matching with the fundamental
(string)
theory is done. Scale independence of the effective action guarantees
independence of the effective theory with respect to the choice of the
scale
$\Lambda$. For practical purposes it is customary to take
$\Lambda=M_{st}$,
where the boundary conditions are provided from the underlying theory.
The two terms inside the bracket in the second line of (\ref{En})
correspond, respectively, to the 1-loop and
the dominant (logarithmic) two-loop contribution due to the
wave-function
renormalization  of the i-th bulk mode, which is coupled to the massless
(twisted) fields in the boundary with coupling $\alpha_i$. This coupling
denotes generically either the gauge or the Yukawa couplings between
fields in the bulk and in the boundary. Since only even fields couple to
the boundary these interactions are $N=1$ supersymmetric. For those
fields
in the bulk  without gauge and Yukawa interactions with the boundary
(as e.g. the gravitational and moduli multiplets) $\alpha_i\equiv 0$.
The dots stand for the remaining subdominant (non logarithmic) two-loop
contribution, as well as for higher loops. Note that the two-loop
contribution of only bulk fields has no logarithmic dependence in $M_c$
in
analogy with finite temperature, while the two-loop contribution of only
boundary fields vanishes due to supersymmetry. $\gamma_i$ is a positive
numerical  coefficient (the eigenvalue of $\gamma_T$ for the i-th bulk
field) coming from the one-loop integration over the boundary states;
its
sign is always positive as boundary fields can never be gauge bosons in
the present context.

By imposing the $\Lambda$-independence of (\ref{En}) we can deduce the
$\beta$-functions for $\eta_i$. To lowest order they are given by:
\be
\beta_i\equiv\Lambda\,\frac{d\eta_i}{d\Lambda}=
4\eta_i\gamma_i\frac{\alpha_i}{2\pi}\, ,
\label{betas}
\ee
whose formal solution can be written as
\be
\eta_i(M_c)=\eta_i(\Lambda)\exp\left\{4\gamma_i\int_0^t
\frac{\alpha_i(t')}{2\pi}dt'
\right\}\, ,
\label{etai}
\ee
where $t\equiv\ln(M_c/\Lambda)$. The logarithmic dependence of the
two-loop vacuum energy and the corresponding $\beta$-functions
(\ref{betas}) follow from the one-loop running of all untwisted masses
$M_i^2=\eta_i^{1/2} M_c^2$ due to the wave function renormalization of
bulk fields from
the massless twisted loops. As a result, the logarithms appearing in the
two-loop expression (\ref{En}) can be absorbed in the (one-loop)
renormalized
masses $M_i(M_c)$.

The two-loop result (\ref{En}) can be resummed to all-loop in the
leading-log approximation by the improved vacuum energy:
\be
E=\eta(M_c)\, M_c^4\, ,
\label{Resum}
\ee
where $\eta(M_c)=\sum_i \eta_i(M_c)$ and $\eta_i(M_c)$ is defined in
(\ref{etai}). Minimization of eq.~(\ref{Resum}) with respect to $M_c$
leads
to
\be
M_c\frac{dE}{dM_c}=4M_c^4\left[\eta(M_c)+\Gamma(M_c)\right]=0\, ,
\label{derivada}
\ee
where $\Gamma=\sum_i\eta_i\gamma_i\alpha_i/2\pi$. The minimum of the
potential
is then given by the value of $M_c$ such that
\be
\eta(M_c)+\Gamma(M_c)=0\; ,
\label{extremal}
\ee
and
\be
\Gamma(M_c)+\frac{1}{4}\, M_c\, \frac{d\Gamma(M_c)}{dM_c}>0\, ,
\label{minimum}
\ee
which is the condition for the extremal (\ref{extremal}) to be a
minimum.

This phenomenon, i.e. the appearance of a minimum by radiative
corrections,
has been long ago known as dimensional transmutation~\cite{cw,nos}, as
one
dimensionless parameter, $\eta$, is traded for the VEV of a field,
$M_c$.
The physical picture by which $M_c$ does acquire a VEV is then similar
to
radiative breaking in field theory. We start running the
$\eta_i$-parameters at the scale $\Lambda=M_{st}$ where the fundamental
theory gives us the boundary values of all couplings $\alpha_i(M_{st})$.
At
the boundary,
$\eta(M_{st})$ and $\Gamma(M_{st})$ should not
satisfy eq.~(\ref{extremal}). As we go down with the energy
the quantity $\eta+\Gamma$ should approach zero and, at a given scale
$M_c$, it should change sign. However, taking into account from
eq.~(\ref{etai}) that $\eta_i(M_c)$ is a monotonically
decreasing (increasing) function for $\eta_i(\Lambda)>0$
($\eta_i(\Lambda)<0$), it follows
that some $\eta_i(M_c)$ are required to be positive and
some $\eta_i(M_c)$ should be negative. On the other hand,
notice that $\eta_i$ are numerical factors depending on the R-charges
used to break supersymmetry. In particular, the contribution
to $\eta$ from a single bosonic and fermionic degree of freedom is given
by:
\be
\Delta\eta=-{3\over 128\pi^6}\left[Li_5(e^{2i\pi q_B})-Li_5(e^{2i\pi
q_F})
+h.c.\right]\, .
\label{c1}
\ee
The expression (\ref{c1}) is negative for $q_B=0$ and
$q_F=1/2$, which means that if the fermion number operator
$(-1)^F$ is used for the Scherk-Schwarz breaking the condition
(\ref{extremal}) for radiative determination of the
compactification radius is never realized. However other
R-symmetries (as e.g. the $SU(2)_R$ of $N=2$ supersymmetry)
might provide different signs for different sectors and yield
the necessary conditions for radiative breaking.

We can now expand eq.~(\ref{extremal}) to lowest order and find an
approximated solution for the non-trivial minimum as
\be
M_c=\exp\left\{-\frac{1}{4}
\left(\frac{\eta(\Lambda)}{\Gamma(\Lambda)}+1+{\cal O}(\hbar)\right)
\right\}\Lambda\, .
\label{mc}
\ee
For $\Lambda=M_c$ the solution (\ref{mc}) satisfies trivially
eq.~(\ref{extremal}) to lowest order,
as it should, while for $\Lambda=M_{st}$, $M_c$
can be hierarchically smaller than the string scale, depending on the
particular string model and on the value of the gauge couplings
$\alpha_i$
at the string (unification) scale. In the latter case the large
logarithm
developed by the minimum does not invalidate
perturbation theory, it just reflects a bad choice of the scale and  can
be
reabsorbed in the renormalized parameters.

A very simple example can be provided by a model where only the
strong coupling is kept and all other couplings $\alpha_i$ (electroweak,
Yukawa, gravitational,...) are neglected. Then we have a strongly
coupled
gauge (gluino vector multiplet) sector with
$\eta_s\equiv\eta_s(M_{st})<0$
and a
non-interacting sector (electroweak vector multiplets, Higgs and
gravitational multiplets,...) with
$\eta_0\equiv\eta_0(M_{st})>0$. The sign of $\eta_s$ is unambiguous
since
for vector multiplets $q_B=0$. The sign and value of $\eta_0$ depends of
course on the field content and the R-invariance used for the
Scherk-Schwarz mechanism. We will consider
$\rho=\eta_0/\eta_s$ as a free parameter and adopt the running of
the strong coupling given in (\ref{run}) and the string scale
and unification coupling obtained in (\ref{mst}).

In the region of scales between $M_{st}$ and $M_c$ the running
of $\alpha_i$ is dominated by the linear term. Neglecting the
logarithmic term in (\ref{run}) we can obtain an analytic
expression for $\eta_s(M_c)$, and so for the vacuum energy as:
\be
\frac{1}{\eta_s}E=\rho-\left[\frac{M_c}{M_{st}}+\frac{b_s^{KK}
\alpha_{st}}
{2\pi}
\left(1-\frac{M_c}{M_{st}}\right)\right]^{4\gamma_s/\left(\frac{2\pi}
{\alpha_{st}}-b_s^{KK}\right)}\, ,
\label{energia}
\ee
where $b_s^{KK}=-6$ and $\gamma_s=6$ is the contribution to the
anomalous dimension of gluons from the chiral quark
supermultiplets in the boundary.
\begin{figure}[htb]
\centerline{
\psfig{figure=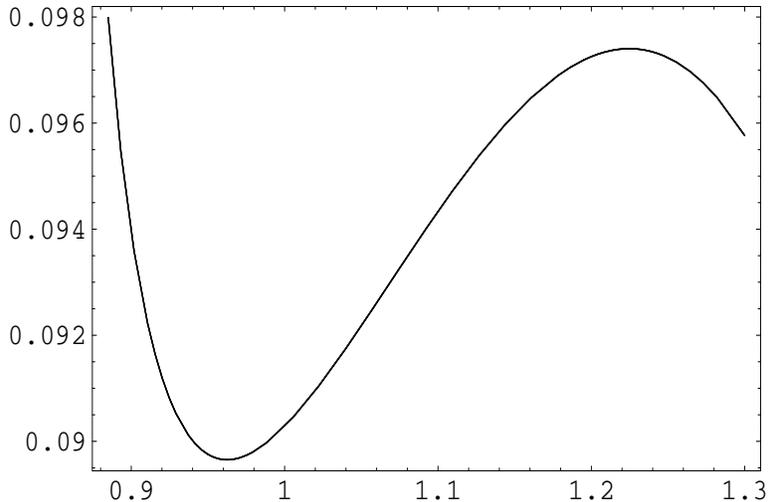,width=10cm,height=9cm,bbllx=3.cm,bblly=20.cm,bburx=13.cm,bbury=28cm}}

\caption{\it Effective potential/{\rm \,(TeV)}$^4$ as a function of
$M_c/\,
{\rm TeV}$ in the simple model above.}
\end{figure}
In Fig.~1 we have plotted the vacuum energy (\ref{energia}) as a
function of $M_c$, in TeV units, for $\rho=0.74$.
We see that a local minimum develops around 1 TeV.

To conclude, this mechanism can therefore be used to fix the size of the
dimension that breaks supersymmetry at a TeV, in either $N=4$ case, with
logarithmic unification of gauge couplings, or in the generic $N=2$ case
with power low evolution and the string scale near the TeV region. In
this
case, however, the generic bulk contribution to the vacuum energy is
much
bigger than $M_c^4$ due to the existence of $n$ additional ultra-large
dimensions of size $r$, that are required to account for the weakness of
four-dimensional gravity:
\be
E_{\rm bulk}\sim M_c^{4+n}r^n\sim M_{st}^2 M_{P\ell}^2\qquad\qquad{\rm
for}\quad
M_c\sim M_{st}\, .
\label{Ebulk}
\ee
The scaling $M_c^{4+n}$ is a consequence of the Scherk-Schwarz breaking
in
$4+n$ non-compact dimensions and can also be understood from the
four-dimensional viewpoint as the multiplicity $(rM_c)^n$ of the
KK-towers
with respect to the $n$ ultra-large dimensions. Since these KK-states
have
no standard model gauge interactions, there are no logarithmic
corrections.
As shown in eq.~(\ref{Ebulk}), this bulk contribution brings back
essentially the problem of quadratic divergences after supersymmetry
breaking and invalidates the radiative determination of the
compactification scale $M_c$. In the context of TeV strings this problem
is
even worse since such a cosmological constant induces a new scale much
bigger than $M_{st}$. One has therefore to impose the condition that
this
bulk contribution to the vacuum energy vanishes. This selects out
special
models having equal number of bosons and fermions in the
$(4+n)$-dimensional bulk after supersymmetry breaking, level by level,
at
least perturbatively \cite{ks}. The next dominant contribution is then
$M_c^4$ up to logarithms and the above mechanism of fixing $M_c$ can be
applied. Of course, the problem of determining the additional
ultra-large
radii $r$ still remains open.

As a result, in both $N=4$ and generic $N=2$ cases, the
compactification scale is determined in terms of the string
scale and the unification coupling, and it can be
hierarchically smaller. It is then remarkable that once the
compactification scale is fixed, the phenomenology of
supersymmetry breaking in both cases is very little distinct,
due to the extreme softness of the Scherk-Schwarz breaking
that leads to a logarithmic sensitivity of the soft terms in
the string scale only at two loops. For example, from
eq.~(\ref{mc}) it follows that starting with a string scale
$M_{st}\sim 10^{16}$ GeV, one obtains a compactification
scale near the TeV region provided that $\Gamma/\eta={\cal
O}(10^{-2})$, which is reasonable since $\Gamma$ is a
two-loop correction while $\eta$ is a one-loop effect.

\section{Concluding remarks}

The first interesting consequence of our analysis is the
pattern of supersymmetry breaking. While gaugino and Higgsino
masses are of the order of the compactification scale, scalar
masses are generated at one loop level via gauge interactions
and are naturally one order of magnitude lighter. Thus,
flavor universality is guaranteed as in gauge mediated
models. On the other hand, again as in gauge mediation, the
stop correction to the Higgs mass-squared drives it to
negative values, breaking the electroweak symmetry. The
resulting spectrum consists of heavy charginos and
neutralinos ($2-3$ TeV), squarks at $400-500$ GeV, and the
right-handed slepton as the lightest supersymmetric particle
with mass close to the electroweak scale. Moreover, the
higher dimensional nature of the theory and the softness of
the Scherk-Schwarz breaking replace effectively the
ultraviolet cutoff with the compactification scale, keeping
the loop corrections to the Higgs mass due to the heavy
gauginos small. Thus, one obtains a pattern of supersymmetry
breaking with hierarchical structure, which is very different
from all other scenaria. In addition, the models we study are
extremely predictive, since they have no free parameters,
other than a discrete option of boundary conditions, and the
superparticle spectrum is fully determined.

Furthermore, this mechanism offers a possibility to determine
dynamically
the compactification scale by relating it to the fundamental (string)
scale
in a hierarchical way. One may think naively that the radius modulus
dependent potential, generated by the vacuum energy, would be runaway
and
the extra dimension either decompactifies, or else, it shrinks to zero
size
in the minimum. However, in the presence of boundary (twisted) fields
with
gauge  interactions, there are logarithmic corrections that can
stabilize
the  radius at a non-trivial minimum. Moreover, its value has an
exponential sensitivity to the coefficient of the one loop logarithm,
and
thus, it may be hierarchically smaller that the string scale. As a
result,
this mechanism can generate a very large or smaller value for the
compactification scale, depending on the detailed spectrum of the model.

{\large\bf Acknowledgements}

It is a pleasure to thank Nima Arkani-Hamed, Antonio Delgado, Jaume Garriga,
Nikolai Krasnikov and Riccardo
Rattazzi for very valuable discussions. AP thanks the
Physics Department of Stanford University, MQ thanks the
IFAE of Universitat Aut\`onoma de Barcelona and the Centre de
Physique Th\'eorique of the Ecole Polytechnique, and all
authors thank the CERN Theory Division for hospitality.
This research is supported in part by the EEC under TMR
contract ERBFMRX-CT96-0090. The work of SD 
 is supported by NSF grant NSF-PHY-9870115.
The work of AP is supported in part
by CICYT under contract AEN95-0882. The work of MQ is supported in part 
by CICYT under contract AEN98-0816. The work of IA and MQ is supported in part
by IN2P3-CICYT contract Pth 96-3.


\newpage
\begin{thebibliography}{99}

\bibitem{dg} S. Dimopoulos and H. Georgi, \NPB{193}{81}{150}.
\bibitem{gr} For a recent review see G. Giudice and R. Rattazzi,
hep-ph/9801071 and references therein.
\bibitem{a} I. Antoniadis, \PLB{246}{90}{377}; Proc.
PASCOS-91 Symposium, Boston 1991 (World Scientific, Singapore)
p.718; I. Antoniadis and K. Benakli, \PLB{326}{94}{69};
I. Antoniadis, K. Benakli and M. Quir\'os, \PLB{331}{94}{313};
K. Benakli, \PLB{386}{96}{106}.
\bibitem{amq} I. Antoniadis, C. Mu\~noz and M. Quir\'os,
\NPB{397}{93}{515}.
\bibitem{add} I. Antoniadis, S. Dimopoulos and G. Dvali,
\NPB{516}{98}{70}.
\bibitem{pq} A. Pomarol and M. Quir\'os, hep-ph/9806263, \PLB{}{98}{} to
appear.
\bibitem{ss} J. Scherk and J.H. Schwarz, \NPB{153}{79}{61} and
\PLB{82}{79}{60};
E. Cremmer, J. Scherk and J.H. Schwarz, \PLB{84}{79}{83};
P. Fayet, \PLB{159}{85}{121} and \NPB{263}{86}{649}.
\bibitem{kp} R. Rohm, \NPB{237}{84}{553};
C. Kounnas and M. Porrati, \NPB{310}{88}{355}; S.
Ferrara, C. Kounnas, M. Porrati and F. Zwirner, \NPB{318}{89}{75};
C. Kounnas and B. Rostand, \NPB{341}{90}{641};  E. Kiritsis, C. Kounnas,
P.M. Petropoulos and J. Rizos, \PLB{385}{96}{87}.
\bibitem{ablt} I. Antoniadis, C. Bachas, D. Lewellen and T. Tomaras, \pl
207
(1988) 441.
\bibitem{aadd} N. Arkani-Hamed, S. Dimopoulos and G. Dvali,
\PLB{429}{98}{263}
and hep-ph/9807344; I.  Antoniadis, N. Arkani-Hamed, S. Dimopoulos and
G. Dvali, \PLB{436}{98}{257}.
\bibitem{l} J. D. Lykken, \PRD{54}{96}{3693}; G. Shiu and S.-H.H. Tye,
hep-th/9805157.
\bibitem{fkz} S. Ferrara, C. Kounnas and F. Zwirner,
\NPB{429}{94}{589}.
\bibitem{susy98} M. Quir\'os, talk given at the SUSY 98 Conference,
Oxford, England, 11-17 July 1998, http://hepnts1.rl.ac.uk/SUSY98/.
\bibitem{ddg} K.R. Dienes, E. Dudas and T. Gherghetta, \PLB{436}{98}{55}
and hep-ph/9806292.
\bibitem{Ross} D. Ghilencea and G.G. Ross, hep-ph/9809217.
\bibitem{prep} A. Delgado, A. Pomarol and M. Quir\'os, in preparation.
\bibitem{casimir} H.B.G. Casimir, {\it Proc. Kon. Nederl. Akad.
Wetensch.}
{\bf 51} (1948) 793.
\bibitem{peskin} E.A. Mirabelli and M. Peskin, \PRD{58}{98}{065002}.
\bibitem{opal} See, for example, The OPAL Collaboration, \PLB
{433}{98}{195}.
\bibitem{derujula} A. De Rujula, S.L. Glashow, U. Sarid,
\NPB{333}{90}{173};  S. Dimopoulos, D. Eichler, R.
Esmailzadeh and G.D. Starkman, \PRD{41}{90}{2388}.
\bibitem{cohen} R.S. Chivukula, A.G. Cohen, S. Dimopoulos and
T.P. Walker \PRL{65}{90}{957}.
\bibitem{gould}A. Gould, B.T. Draine, R.W. Romani and
S. Nussinov, \PLB{238}{90}{337}.
\bibitem{aleph} See, for example, The ALEPH Collaboration,
{\it Eur. Phys. J.} {\bf C4} (1998) 433.
\bibitem{k} N.V. Krasnikov, {\em Pis'ma v ZhETF} {\bf 67} (1998) 727.
\bibitem{dgp} G. Dvali, G.F. Giudice and A. Pomarol,
\NPB{478}{96}{31}.
\bibitem{cw} S. Coleman and E. Weinberg, \PRD{7}{73}{1888}.
\bibitem{nos} C. Kounnas, F. Zwirner and I. Pavel, \PLB{335}{94}{403}
and references therein.
\bibitem{ks} S. Kachru, J. Kumar and E. Silverstein, hep-th/9807076;
J.A. Harvey, hep-th/9807213; G. Shiu and S.-H.H. Tye, hep-th/9808095;
S. Kachru, and E. Silverstein, hep-th/9810129.

\end{thebibliography}
\end{document}